\documentclass[twocolumn,showpacs,preprintnumbers,amsmath,amssymb]{revtex4-1}
\usepackage{hyperref}
\usepackage{graphicx}
\usepackage{tikz}

\begin{document}

\title{Lensing in the McVittie metric}
\author{Oliver F. Piattella}
\email{of.piattella@uninsubria.it}
\affiliation{Physics Department, Universidade Federal do Esp\'irito Santo, Vit\'oria, 29075-910, Brazil}

\date{\today}


\begin{abstract}
 We investigate the effect of the cosmological expansion on the bending of light due to an isolated point-like mass. We adopt McVittie metric as the model for the geometry of the lens. Assuming a constant Hubble factor we find an analytic expression involving the bending angle, which turns out to be unaffected by the cosmological expansion at the leading order.
 \end{abstract}


\maketitle

\section{Introduction}

McVittie metric \cite{McVittie:1933zz} is a spherically symmetric solution of Einstein's equations which asymptotically tends to a Friedmann-Lema\^itre-Robertson-Walker (FRLW) universe. It was introduced in 1933 by McVittie in order to investigate cosmological effects on local systems, e.g. on closed orbits of planets or stars. This issue has been examined again in the past years \cite{1996MNRAS.282.1467B, Faraoni:2007es, Carrera:2008pi, Bochicchio:2012vu, Kopeikin:2012by, Nolan:2014maa}, especially in relation to the Pioneer anomaly \cite{Anderson:2001ks}. It is still a matter of debate whether and how much cosmological effects influence the physics of local systems.

McVittie metric has been intensively analyzed by many authors, see e.g. Refs.~\cite{Nolan:1998xs, Nolan:1999kk, Nolan:1999wf, Kaloper:2010ec, Lake:2011ni, Nandra:2011ug, Nandra:2011ui}. In particular, Nolan analyzed the mathematical properties of McVittie solution in a series of three papers \cite{Nolan:1998xs, Nolan:1999kk, Nolan:1999wf}. One of the most important results is that McVittie metric is not a black hole solution because where one expects a horizon, there is instead a weak singularity (i.e. geodesics can be extended through it). There is an exception to this theorem: when the external, FLRW part of McVittie solution tends to be cosmological constant-dominated \cite{Kaloper:2010ec, Lake:2011ni}. 

Taking advantage of the well-posedness of McVittie metric with flat spatial hypersurfaces \cite{Nolan:1998xs} (see also Ref.~\cite{Nandra:2011ug}), we use this solution in order to understand how the deflection of light caused by a point mass is affected by the embedding of the latter in an expanding universe. This idea has been recently explored in Ref.~\cite{Aghili:2014aga}, where the authors numerically show that an effect due to the Hubble constant $H_0$ does exist on the deflection angle. 

There is an ample literature on this problem, mostly specialized to the case in which a cosmological constant $\Lambda$ dominates and, for this reason, based on Kottler (Schwarzschild-de Sitter) metric \cite{kottler1918physikalischen}. In his pioneering investigation, Islam \cite{islam1983cosmological} found no influence whatsoever by $\Lambda$ on the bending of light. Only less than a decade ago, Ishak and Rindler \cite{Rindler:2007zz, Ishak:2008zc}, via a new definition of the bending angle, showed that an effect due to $\Lambda$ seems to exist. Their work and results gave rise to many others investigations, see e.g. Refs.~\cite{Schucker:2007ut, Ishak:2008ex, Park:2008ih, Sereno:2008kk, Simpson:2008jf, Khriplovich:2008ij, Ishak:2010zh, Biressa:2011vy, Hammad:2013wda, Butcher:2016yrs}. There seems to be common agreement now that $\Lambda$ indeed affects the bending of light. A debate actually exists on the entity of this influence. Among the works cited above, Refs.~\cite{Park:2008ih, Simpson:2008jf, Khriplovich:2008ij, Butcher:2016yrs} disagree with the existence of any relevant effect caused by $\Lambda$ on the lensing phenomenon. The reason is essentially the following: putting source, lens and observer in a cosmological setting, i.e. taking into account the Hubble flux, makes the $\Lambda$ contribution completely negligible (but non zero in principle) because of how it enters the definition of the angular diameter distances and because of aberration effects due to the relative motion.

In order to better understand these points, we present here a perturbative, analytic calculation of the bending angle in McVittie metric. We assume a constant Hubble parameter, thereby focusing on the case of a cosmological constant-dominated universe. By using McVittie metric in the coordinates of Eq.~\eqref{mcvittiecomoving}, we take into account the embedding of source, lens and observer in a cosmological context, thereby potentially addressing the issues raised in Refs.~\cite{Park:2008ih, Simpson:2008jf, Khriplovich:2008ij, Butcher:2016yrs}. 

We find no extra contribution to the bending angle coming from cosmology. Therefore, we corroborate the results of Refs.~\cite{Park:2008ih, Simpson:2008jf, Khriplovich:2008ij, Butcher:2016yrs}.


The paper is structured as follows. In Sec.~\ref{Sec:McVittiemetric} we present McVittie metric and tackle the lensing problem, calculating the bending angle. In Sec.~\ref{Sec:Alignment} we focus on the case of Einstein's ring systems. Finally, Sec.~\ref{Sec:DiscandConcl} is devoted to discussion and conclusion. We use natural $G = c = 1$ units throughout the paper.

\section{McVittie metric and lensing}\label{Sec:McVittiemetric}

McVittie metric \cite{McVittie:1933zz} has the following form:
\begin{equation}\label{mcvittiecomoving}
	ds^2 = -\left(\frac{1 - \mu}{1 + \mu}\right)^2dt^2 + (1 + \mu)^4a(t)^2(d\rho^2 + \rho^2d\Omega^2)\;,
\end{equation}
where $a(t)$ is the scale factor and
\begin{equation}\label{mudefinition}
	\mu \equiv \frac{M}{2a(t)\rho}\;,
\end{equation}
where $M$ is the mass of the point-like lens. When $\mu \ll 1$, metric~\eqref{mcvittiecomoving} can be approximated by
\begin{equation}\label{mcvittiepert}
	ds^2 = -\left(1 - 4\mu\right)dt^2 + (1 + 4\mu)a(t)^2(d\rho^2 + \rho^2d\Omega^2)\;,
\end{equation}
which is the usual perturbed FLRW metric in the Newtonian gauge and $2\mu$ is what is usually called gravitational potential. 

\begin{figure}[h!]
	\begin{tikzpicture}
	\draw (0,0) -- (7,0);
	\draw (0,0) -- (0,1);
	\draw (0,0.5) node[right] {$b$};
	\fill (7,0) circle (2pt) node[below right] {Us};
	\fill (3.5,0) circle (2pt) node[above] {Lens};
	\fill (0,1) circle (2pt) node[above] {Source};
	\draw[line width=2pt] (0, 1) .. controls(3.5,0.7) .. (7, 0);
	\draw[|<->|] (0,-0.5) -- (7,-0.5);
	\draw[|<->|] (3.5,-1) -- (7,-1);
	\draw (2,-0.5) node[below] {$\chi_S$};
	\draw (5,-1) node[below] {$\chi_L$};
	\draw (4,0) -- (4,0.7);
	\draw[|<->|] (4,0.7) -- (7,0.7);
	\draw (5,0.7) node[above] {$\chi(t)$};
\end{tikzpicture}
\caption{Scheme of lensing.}
\label{figuretrajectory}
\end{figure}
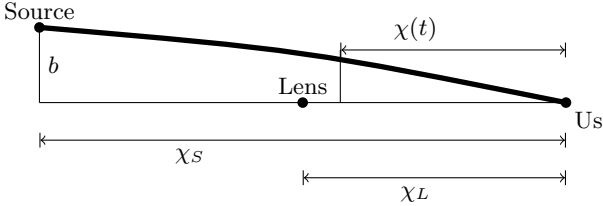

We adopt the same formalism Dodelson uses in Chapter 10 of his textbook \cite{Dodelson:2003ft}. We use as time-variable the background comoving distance $\chi$ (as if it there were no point-like mass) from us to the plane where the photon is at a certain time $t$. See Fig.~\ref{figuretrajectory}. 

The relation between $\chi$ and the background expansion is the usual one for the FLRW metric:
\begin{equation}
	\frac{d\chi}{dt} = -\frac{1}{a}\;,
\end{equation}
and the comoving distances of the source and of the lens, $\chi_S$ and $\chi_L$ respectively, do not change.

The null geodesics equation for the transversal displacement $l^i$ of the photon is the following:
\begin{eqnarray}\label{transvdisplac}
	\frac{a}{p}\frac{1-\mu}{1+\mu}\frac{d}{d\chi}\left(\frac{p}{a}\frac{1+\mu}{1-\mu}\frac{dl^i}{d\chi}\right) = \frac{2(1-\mu)}{(1+\mu)^7}\delta^{il}\partial_l\mu\nonumber\\ + 2Ha\left[1 + \frac{2\partial_t\mu}{(1+\mu)H}\right]\frac{dl^i}{d\chi}\nonumber\\ - \frac{2}{1+\mu}\left(\delta^i_j\partial_k\mu + \delta^i_k\partial_j\mu - \delta_{jk}\delta^{il}\partial_l\mu\right)\frac{dl^j}{d\chi}\frac{dl^k}{d\chi}\;,
\end{eqnarray}
where $i = 1,2$; $H \equiv \partial_t a/a$ is the Hubble factor and $p$ is the photon proper momentum. Since McVittie metric is spherically symmetric, the two equations for $i = 1,2$ are identical. 


Now we do consider $\mu$ small and investigate how a point mass in the expanding universe affects the trajectory of a light ray. In order to do this, in Eq.~\eqref{transvdisplac} we consider $\mu \ll 1$ and small displacements $l^i \ll \chi$. 



Equation~\eqref{transvdisplac} can be then simplified as follows:
\begin{equation}
	\frac{d^2l^i}{d\chi^2} = 4\partial_i\mu\;.
\end{equation}
Using Eq.~\eqref{mudefinition} in the equation above, one gets:
\begin{equation}\label{xthetaeq}
	\frac{d^2l}{d\chi^2} = -\frac{2Ml}{a(\chi)\left[(\chi - \chi_L)^2 + l^2\right]^{3/2}}\;,
\end{equation}
 where we dropped the index $i$ denoting the transversal direction, thanks to spherical symmetry. With the following definitions:
\begin{equation}
	x \equiv \frac{\chi}{\chi_L}\;, \qquad \alpha \equiv \frac{2M}{\chi_L}\;, \qquad y \equiv \frac{l}{\chi_L}\;,
\end{equation}
Eq.~\eqref{xthetaeq} becomes:
\begin{equation}\label{fundeqy}
	\frac{d^2y}{dx^2} = -\alpha\frac{y}{a(x)\left[(x - 1)^2 + y^2\right]^{3/2}}\;.
\end{equation}
Note that a vanishing $\alpha$ implies that $a(x)$ has no effect on the trajectory. This is a sort of ``casting out nines", since indeed we do not expect lensing caused by cosmology only.

We suppose the source to be at a comoving distance $\chi_S$ and solve the above equation considering small $\alpha$, via the following expansion of the solution:
\begin{equation}
	y = y^{(0)} + \alpha y^{(1)} + \alpha^2 y^{(2)} + \dots\;,
\end{equation}
and retaining the first order only in $\alpha$. As initial conditions, we choose:
\begin{equation}\label{icfull}
	y(x_S) = y_S\;, \qquad y(0) = 0\;,
\end{equation}
which mean that the light ray starts from the source with an impact parameter $b \equiv y_S\chi_L$, see Fig.~\ref{figuretrajectory}, and it must arrive to us in order to be detected, thus $y(0) = 0$.

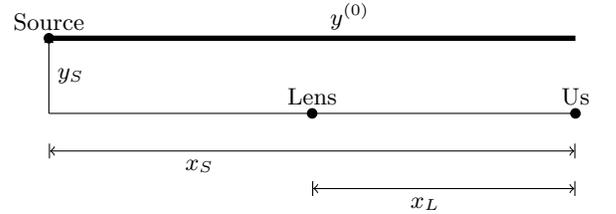
\begin{figure}[h!]
	\begin{tikzpicture}
	\draw (0,0) -- (7,0);
	\fill (7,0) circle (2pt) node[above] {Us};
	\fill (3.5,0) circle (2pt) node[above] {Lens};
	\fill (0,1) circle (2pt) node[above] {Source};
	\draw[line width=2pt] (0, 1) -- (7, 1);
	\draw[|<->|] (0,-0.5) -- (7,-0.5);
	\draw[|<->|] (3.5,-1) -- (7,-1);
	\draw (2,-0.5) node[below] {$x_S$};
	\draw (5,-1) node[below] {$x_L$};
	\draw (4,1) node[above] {$y^{(0)}$};
	\draw[->] (0,0) -- (0,1);
	\draw (0,0.5) node[right] {$y_S$};
\end{tikzpicture}
\caption{Zero-order trajectory.}
\label{figuretrajectoryzerothorder}
\end{figure}

The zero-order solution is trivial, i.e. a straight line:
\begin{equation}
	y^{(0)} = C_1x + C_2\;.
\end{equation}
We choose the two integration constants so that $y^{(0)} = y_S$, i.e. the trajectory is a straight, horizontal line, see Fig.~\ref{figuretrajectoryzerothorder}. 

The first-order solution is then given by the following equation:
\begin{equation}\label{y1eq}
	\frac{d^2y^{(1)}}{dx^2} = -\frac{y_S}{a(x)\left[(x - 1)^2 + y_S^2\right]^{3/2}}\;,
\end{equation}
for which we must choose the following initial conditions:
\begin{equation}
	y^{(1)}(x_S) = 0\;, \qquad y^{(1)}(0) = -y_S/\alpha\;,
\end{equation}
in order to respect those in Eq.~\eqref{icfull} for the full solution.


Let's now consider the contribution of $a(x)$. For a constant Hubble factor $H = H_0$, one can easily determine the scale factor as function of the comoving distance:
\begin{equation}\label{achiHconstant}
	\chi = \int_0^z\frac{dz'}{H(z')} = \frac{z}{H_0} \equiv \frac{1}{H_0}\left(\frac{1}{a} - 1\right)\;, 
\end{equation}
where we solved the integral by introducing the redshift $z$, defined as in the last equality of the above equation. This is, in principle, incorrect because one should also take into account the gravitational redshift caused by the point mass. However, that would produce a second order contribution in $\alpha$ in Eq.~\eqref{fundeqy}, so we neglect it.

Using Eq.~\eqref{achiHconstant}, Eq.~\eqref{y1eq} becomes:
\begin{equation}\label{y1solH0}
	\frac{d^2y^{(1)}}{dx^2} = -\frac{y_S\left(1 + H_0\chi_L x\right)}{\left[(x - 1)^2 + y_S^2\right]^{3/2}}\;.
\end{equation}
In the limit $y_S \ll 1$ the deviation angle 
\begin{equation}
	\delta \equiv \left.\frac{dy}{dx}\right|_{x = 0} - \left.\frac{dy}{dx}\right|_{x = x_S}\;,
\end{equation}
derived by using the solution of Eq.~\eqref{y1solH0} is the following:
\begin{equation}
	\delta = \frac{2\alpha(1 + \chi_L H_0)}{y_S} + \mathcal{O}(y_S)\;.
\end{equation}
Note that Eq.~\eqref{y1solH0} can be solved exactly, but its solution is quite cumbersome so we do not write it down here explicitly. 

Recalling that $\alpha \equiv 2M/\chi_L$ and $y_S = b/\chi_L$, we can write the above formula as:
\begin{equation}\label{corrangle}
	\delta = \frac{4M(1 + \chi_L H_0)}{b} + \mathcal{O}(b/\chi_L)\;.
\end{equation}
This result is similar to the one in the Schwarzschild case, except for the fact that the mass seems to be increased by a relative amount of $H_0\chi_L$ and $b$ is not the proper closest approach distance to the lens, but it is the comoving transversal position of the source. 

From Eq.~\eqref{achiHconstant} we know that $H_0\chi_L = z_L$, i.e. the redshift of the lens. 
Considering the standard $\Lambda$CDM model Friedmann equation
\begin{equation}
	\frac{H^2}{H_0^2} = \Omega_\Lambda + \Omega_{\rm m}(1 + z)^3\;,
\end{equation}
$H$ is approximately constant only as long as $\Omega_\Lambda \gg \Omega_{\rm m}(1 + z)^3$. Using the observed values for the density parameters, approximately $\Omega_\Lambda = 0.7$ and $\Omega_{\rm m} = 0.3$, the above condition amounts to state that $z \ll 0.3$. Therefore, the solution we found in Eq.~\eqref{corrangle} for the bending angle is reliable only when the redshifts involved are very small, much less than 0.3.

Let's consider now the lens equation in the thin lens approximation:
\begin{equation}\label{thinlenseq}
	\theta \approx \beta + \delta\frac{D_{LS}}{D_{S}}\;,
\end{equation}
where $\theta$ is the angular apparent position of the source, $\beta$ is the actual angular position of the source, $D_{\rm LS}$ is the angular distance between the lens and the source and $D_S$ is the angular distance between us and the source. Note that, being metric~\eqref{mcvittiecomoving} in an isotropic form, coordinate angles are equal to physical angles. This can be checked, for example, using the definition introduced in Ref.~\cite{Rindler:2007zz} or via the construction used in Ref.~\cite{Schucker:2007ut}.

Using Eq.~\eqref{achiHconstant}, the angular diameter distance between the source and us can be expressed as
\begin{equation}\label{DSdistance}
	D_S = \frac{1}{H_0}(1 - a_S) = \frac{1}{H_0}\frac{z_S}{1 + z_S}\;.
\end{equation}
The angular diameter distance between the lens and us has a similar form:
\begin{equation}\label{DLdistance}
	D_L = \frac{1}{H_0}(1 - a_L) = \frac{1}{H_0}\frac{z_L}{1 + z_L}\;.
\end{equation}
On the other hand, $D_{LS} \neq D_S - D_L$, but, see e.g. Ref.~\cite{Peacock:1999ye}:
\begin{equation}\label{DLSdistance}
	D_{LS} = a_S(\chi_{S} - \chi_L) = \frac{1}{H_0}\frac{z_S - z_L}{1 + z_S}\;.
\end{equation}
Using Eq.~\eqref{corrangle}, Eq.~\eqref{thinlenseq} can be written as follows:
\begin{equation}\label{newlensequation}
	\theta \approx \beta + \frac{4M(1 + z_L)}{b}\frac{D_{LS}}{D_S}\;.
\end{equation}
For comparison, let's consider the standard lens equation, which is based on the union of the results coming from Schwarzschild metric and from FLRW metric, see e.g. Refs.~\cite{Peacock:1999ye, Weinberg:2008zzc}: 
\begin{equation}\label{Weinbergformula}
	\theta \approx \beta + \frac{4M}{r_0}\frac{D_{LS}}{D_S}\;.
\end{equation}
Here, $r_0$ is the closest approach distance to the lens. In a cosmological setting, it is considered as a proper distance and, therefore, expressed as $r_0 = \theta D_L$. In our Eq.~\eqref{newlensequation}, $b$ is the comoving transversal position of the source, thus apparently we have a different result. However, this may not be the case because in comoving coordinates $b$ can also be interpreted as the impact parameter and as the closest approach distance to the lens. Indeed, $y^{(0)} = y_S$ is the zero-order solution and any correction to it is a first-order quantity in $\alpha$ which, if substituted into Eq.~\eqref{newlensequation}, would give a sub-dominant contribution.

Therefore, at the leading order, we can write $b = \theta\chi_L$. The factor $(1 + z_L)$ of Eq.~\eqref{newlensequation} combined with $\chi_L$ gives precisely the angular diameter distance to the lens: $b/(1 + z_L) = \theta\chi_L/(1 + z_L) = \theta D_L$. No difference from the standard case is found.


\section{In case of alignment}\label{Sec:Alignment}

\begin{figure}[h!]
	\begin{tikzpicture}
	\draw (0,0) -- (7,0);
	\fill (7,0) circle (2pt) node[above] {Us};
	\fill (4,0) circle (2pt) node[below] {Lens};
	\fill (0,0) circle (2pt); 
	\draw (0,0.2) node[above] {Source};
	\draw (1.5,0) arc (0:20:1);
	\draw (1.75,-0.05) node[above] {$\theta_S$};
	\draw (6,0) arc (180:160:1);
	\draw (5.7,-0.05) node[above] {$\theta_E$};
	\draw[line width=2pt] (0, 0) .. controls(4, 1) .. (7, 0);
	\draw[|<->|] (0,-0.5) -- (7,-0.5);
	\draw[|<->|] (4,-1) -- (7,-1);
	\draw (2,-0.5) node[below] {$\chi_S$};
	\draw (5,-1) node[below] {$\chi_L$};
	\draw (5,0) -- (5,0.7);
	\draw[|<->|] (5,0.7) -- (7,0.7);
	\draw (6,0.7) node[above] {$\chi$};
	\draw (4,0) -- (4,0.7);
	\draw (4,0.3) node[left] {$b$};
	\end{tikzpicture}
\caption{Scheme of lensing in case of alignment.}
\label{figuretrajectoryeinsteinring}
\end{figure}
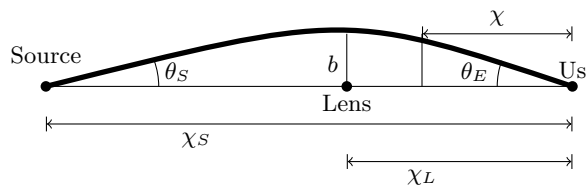

The solution we have exploited in the previous section does not work in the case of alignment among source, lens and observer. See Fig.~\ref{figuretrajectoryeinsteinring}. 

In this case, the zero order solution cannot be a horizontal trajectory, because it would never reach us. The zero order trajectory is now
\begin{equation}
	y^{(0)} = \theta_S(x_S - x)\;,
\end{equation}
where we assume $\theta_S \ll 1$. In order for the trajectory to reach us, we must choose the initial condition $y^{(1)}(0) = -\theta_Sx_S/\alpha$. 

Computing again the deflection angle, we get:
\begin{equation}\label{bendangEinst}
	\delta = \frac{4M(1 + \chi_L H_0)}{\theta_S(\chi_S - \chi_L)} + \frac{M\mathcal{O}(\theta_S)}{\chi_S - \chi_L}\;.
\end{equation}
But, the simple geometry of the lensing process in Fig.~\ref{figuretrajectoryeinsteinring} shows that:
\begin{equation}
	\theta_S(\chi_S - \chi_L) = b = \theta_E\chi_L\;,
\end{equation}
and $\chi_L/(1 + z_L) = D_L$, reproducing thus the standard result. Again, at the leading order no cosmological correction appears.

\section{Discussion and conclusions.}\label{Sec:DiscandConcl}

We have analyzed the issue of whether cosmology affects local phenomena such as the bending of light by a compact mass. To this purpose, we adopted McVittie metric, which describes the geometry of a point mass (the lens, in our picture) in the expanding universe. We assume a constant Hubble factor, thereby presupposing a cosmological constant-dominated FLRW universe surrounding the point mass. We found no correction coming from cosmology at the leading order in the bending angle, thus corroborating the results of \cite{Park:2008ih, Simpson:2008jf, Khriplovich:2008ij, Butcher:2016yrs}.

\begin{acknowledgements}
The author is grateful to L. Bombelli, A. Maciel da Silva, V. Marra, D. C. Rodrigues, T. Sch\"ucker, and the anonymous referees for important suggestions and stimulating discussions. The author also thanks CNPq (Brazil) for partial financial support.
\end{acknowledgements}

\bibliography{McVittie}

\end{document}